# Labeling subtypes in a Parkinson's Cohort using Multifeatures in MRI : Integrating Grey and White Matter Information

Tanmayee Samantaray[1*], Jitender Saini[2], Pramod Kumar Pal[3], Bithiah Grace Jaganathan[4,5], Vijaya V Saradhi[6] and Gupta CN[1]

[1]Neural Engineering Lab, Department of Biosciences and Bioengineering, Indian Institute of Technology Guwahati, Guwahati, Assam, 781039, India. Email: tanma176106113@iitg.ac.in, cngupta@iitg.ac.in

[2]Department of Neuroimaging and Interventional Radiology, National Institute of Mental Health and Neurosciences, Bengaluru, 560029, India. Email: jsaini76@gmail.com

[3]Department of Neurology, National Institute of Mental Health & Neuro Sciences, Bengaluru, 560029, India. Email: palpramod@hotmail.com

[4]Stem Cells and Cancer Biology Research Group, Department of Biosciences and Bioengineering, Indian Institute of Technology Guwahati, Guwahati, Assam, 781039, India. Email: bithiahgj@iitg.ac.in

[5]Jyoti and Bhupat Mehta School of Health Sciences and Technology, Indian Institute of Technology Guwahati, Guwahati, Assam, 781039, India.

[6]CSE Department, Indian Institute of Technology Guwahati, Guwahati, Assam, 781039, India. Email: saradhi@iitg.ac.in

*Corresponding author: Tanmayee Samantaray, Email: tanma176106113@iitg.ac.in

<u>Highlights</u>

- Joint independent component analysis (ICA) based data mining approach using grey and white matter stratified Parkinson's disease into three distinct subtypes- A, B and AB.
- Subtype B is the mild-severe type with mild motor symptoms.
- Subtype A is identified as the intermediate type with clinical scores higher than subtype B and lower than subtype AB.
- Subtype AB is the most-severe type with predominance in motor impairment.
- Mutual K Nearest Neighbor-based thresholding resulted to binarization of weighted matrix for network analysis.

| ARTICLE INFO | ABSTRACT |
|---|---|
|  | Thresholding of networks has long posed a challenge in brain connectivity analysis. Weighted networks are typically binarized using threshold measures to facilitate network analysis. Previous studies on MRI-based brain networks have predominantly utilized density or sparsity-based thresholding techniques, optimized within specific ranges derived from network metrics such as path length, clustering coefficient, and small-world index. Thus, determination of a single threshold value for facilitating comparative analysis of networks remains elusive. To address this, our study introduces Mutual K Nearest Neighbor (MKNN)-based thresholding for brain network analysis. Here, nearest neighbor selection is based on the highest correlation between features of brain regions.<br>Construction of brain networks was accomplished by computing Pearson correlations between grey matter volume and white matter volume for each pair of brain regions. Structural MRI data from 180 Parkinson's disease (PD) patients and 70 healthy controls from the National Institute of Mental Health and Neurosciences, India were analyzed. Additionally, subtypes within Parkinson's disease were identified based on grey and white matter volume atrophy using source-based morphometric decomposition. The loading coefficients were correlated with clinical features to discern clinical relationship with the deciphered subtypes.<br>Our data-mining approach revealed three distinct subtypes: Subtype A (N = 51, intermediate type with clinical scores between subtypes B and AB), Subtype B (N = 57, mild-severe type with mild motor symptoms), and Subtype AB (N = 36, most-severe type with predominance in motor impairment). Subtype-specific weighted matrices were binarized using MKNN-based thresholding for brain network analysis. Permutation tests on network metrics of resulting bipartite graphs demonstrated significant group differences in betweenness centrality and participation coefficient. The identified hubs were specific to each subtype, with some hubs conserved across different subtypes.<br>Clinical Relevance: Treatments tailored to the individual subtype of Parkinson's disease may be investigated using structural brain connections. |

## 1. Introduction

Brain connectivity, as revealed through network analysis techniques, elucidates the interactive functionality of distinct brain regions in information processing (Bassett &

Sporns, 2017). Such analyses offer insights into both structural and functional characteristics of these regions, providing valuable understanding of complex neurological phenomena. In this context, structural magnetic resonance imaging (sMRI) examines the intricate topology of brain networks. Brain networks are usually constructed based on association matrices, which are often binarized across a range (Samantaray et al., 2022b) of thresholds for computation and analysis of network metrics. Some of the threshold methods are density (X. Xu et al., 2018; Yadav et al., 2016), sparsity (Samantaray et al., 2022a, 2023, 2024; Wu et al., 2018) and objective function (Theis et al., 2023). Aggressive thresholding may yield a sparsely connected graph that is unusable, while refraining from thresholding entirely may lead to the presence of numerous false positive edges (Theis et al., 2023). This encourages the need for a single threshold measure. This is highly pertinent as applied to neurodegenerative diseases like Parkinson's disease (PD), where abnormal patterns can be studied.

Mutual K-nearest neighbor (MKNN) is a varied form of KNN that has been widely used in classification studies. The concept of clustering (Jarvis & Patrick, 1973) used shared similarity between two near neighbors based on Euclidean distance. Accordingly, two data points were declared similar when their corresponding K-nearest neighbor lists match. Another study (Hu & Bhatnagar, 2012) intended to find clusters of data points based on inter-point affinities identified by density of the data points. It was based on the idea of two-way affinity of a data point with other points in its vicinity. Graph clustering was also attempted (Sardana & Bhatnagar, 2014) using MKNN approach based on network density as well as edge weights in the graph, thereby finding both cliques and non-clique subnetworks. Interestingly, MKNN concept may be used as a thresholding approach to find connectedness between nodes (brain regions) of a network based on mutual strength of connection. One of the main advantages of this technique is its ability to find densely connected regions irrespective of it being a clique or non- clique.

Brain connectivity in PD, as analyzed using GM volume from sMRI scans, suggests network degeneration with aging (Samantaray et al., 2023). Studies also demonstrate significantly lower network metrics such as clustering coefficient and local efficiency than those of age and gender matched healthy controls (Samantaray et al., 2022a). Changes in structural network of patients with hemi-parkinsonism was explored to observe crucial changes in clustering coefficient and local efficiency, that could serve as a potential biomarker for the disease (X. Xu et al., 2018). Lesser connectivity strength was found (Pereira et al., 2015) in brain regions of PD patients with mild cognitive impairment (PD-MCI) in comparison to healthy control (HC). Additionally, PD-MCI was found to have lower global efficiency but larger characteristic path length in frontal and parietal regions compared to PD with normal cognition (PD-NC). In contrast to HC, this investigation revealed aberrant network structure in the patients. When PD patients with tremors were compared with HC (Zhang et al., 2015), it was shown that some regions had higher network efficiency. The compensating mechanism in PD patients with tremors was supported by a consistently higher global efficiency. Male patients had weaker connection strength (Yadav et al., 2016). These studies have correlated specific measure (eg. GM volume or cortical thickness) between different brain regions.

Parkinson's disease (PD) exhibits heterogeneity in clinical symptoms, reflected by variability in underlying causes, progression and phenotypes among patients surviving with the disease. Although the actual cause behind the disease is still under investigation, neuronal loss in different brain regions is considered amongst them. This usually leads to tissue atrophy in multiple brain regions, associated with region-specific functional loss . Differences in regional atrophy may result to variability between the PD patients, indicating presence of subtypes within the disease. Hence, diagnosing and managing the disease often require a more specific approach to address each subtype's unique spectrum of neuroimaging and clinical features.

Lack of an effective clinical diagnostic scale and absence of a consistent biomarker motivates future research to explore stable PD subtypes (Berg et al., 2014; Samantaray et al., 2021, 2022b; Schalkamp et al., 2022). Studies discovering subtypes in PD using clinical phenotypes only (Eglit et al., 2021; Fereshtehnejad et al., 2017; Rodriguez-Sanchez et al., 2021) encourage additional use of brain imaging data for differential diagnosis (Huppertz et al., 2016; Salvatore et al., 2014). Classification of PD and atypical parkinsonism (Park et al., 2020) and subtyping of PD (Albrecht et al., 2022; Cao et al., 2022) were assessed using multimodal neuroimaging data, usually structural and functional MRI. Among other neuroimaging modality, sMRI is non-invasive and offers visualization of high resolution anatomy of brain tissues, making it suitable for use in various studies (Khanna et al., 2023; Rana et al., 2015, 2017). Structural MRI also enables to identify differences in brain volume between various groups, thereby assisting in disease diagnosis.

Literatures suggest various studies using grey matter (GM) (Rana et al., 2017; Samantaray et al., 2024) and white matter (WM) information for diagnosis and classification of PD from healthy individuals (Liu et al., 2016; Song et al., 2021), overlooking the inter-tissue relationship. On the other hand, studies have shown morphometric changes in one tissue affects the other, suggesting simultaneous consideration of GM and WM information (Pfefferbaum et al., 1994; Rana et al., 2017). Moreover, PD is associated with both grey and white matter differences (Rana et al., 2017; Samantaray et al., 2022b), which are also used in multimodal fusion techniques such as joint-SBM (Guo et al., 2012a; L. Xu, Pearlson, et al., 2009) and joint ICA (Guo et al., 2012a).

Joint ICA (JICA) is a multivariate multimodal technique which was applied in previous studies to find joint sources showing group differences (Guo et al., 2012a; L. Xu, Pearlson, et al., 2009). JICA is a data-driven decomposition method that makes it possible to examine a common mixing matrix obtained from at least two features across subjects (Sui et al., 2023). It uses source-based morphometry (SBM) (L. Xu, Groth, et al., 2009) which employs independent component analysis (ICA) technique that enables identification of spatially covarying patterns. The input information matrix (eg. subject $*$ voxel matrix) is decomposed into independent components (brain regions) containing each subject's tissue information in each region. JICA may combine either structure-structure, structure-function, function-function, brain imaging-genotype or structural MRI-single nucleotide polymorphism data (Sui et al., 2012). Joint ICA was used on both GM and WM on sMRI data to find differences in Alzheimer's patients and controls (Guo et al., 2012a), and found three joint sources where patients had reductions in both GM and WM compared to that of HC. Another study on joint-SBM (L. Xu, Pearlson, et al., 2009) aimed to incorporate multiple tissue types from sMRI and decompose two different sources (gray matter and white matter images) to understand group level differences in the obtained sources. Schizophrenic patients were observed with lower GM and WM than controls, while two other joint sources with higher GM and WM than controls using two joint sources. The concept aimed to establish that GM regions exhibited similar intersubject covariation to that observed in WM regions.

Following are the objectives of the present study:

a. Investigate subtypes within Parkinson's disease using sMRI data using interrelationship between GM and WM.

b. Construction of brain network for the derived subtypes using fused information from both tissue types.

c. Network analysis is performed on adjacency matrix binarized using mutual K-nearest neighbor-based thresholding.

To the best of our knowledge, this is the first study employing MKNN for thresholding brain network.

## 2. Methodology

The outlined methodology for PD subtyping through the utilization sMRI scans encompassed the following primary phases: (1) detecting brain components manifesting differences in grey and white matter between PD patients and HC, (2) selecting specific components with highest group differences, (3) the identification of subtypes among PD patients, achieved by fulfillment of selection criteria based on these components. This was subsequently followed by a comprehensive structural brain connectivity analysis conducted on subjects corresponding to each identified subtype.

### 2.1. Participants Demographics

To undertake this research endeavor, the imaging and clinicopathological information of the 70 healthy controls (HC) and 180 PD patients were sourced from Department of Neurology, National Institute of Mental Health and Neurosciences (NIMHANS), Bangalore, India. Essential demographic details of these participants, including clinical information such as age, gender, Unified Parkinson's Disease Rating Scale at on (UPDRS-On) and off (UPDRS-Off) state, Hoehn and Yahr scale (H&Y), age at onset are presented in Table 1. In situations where clinical variables within the dataset exhibited missing records, imputation was carried out by using winsorized mean based on the available data points pertaining to the respective variables. A 90% winsorization method was adopted in which 10% data was modified (i.e. 5% from each tail). The data was obtained in compliance with NIMHANS Institutional Ethics Committee.

**Table 1.** Demographics of participants from NIMHANS

| | **Healthy Control** | **Parkinson's Disease** | **$\chi^2$ or t-value** | **p-value** |
|---|---|---|---|---|
| Count | 70 | 180 | - | - |
| Age in years (Mean ± SD) | 49.24 ± 10.99 | 54.84 ± 9.78 | -3.92 | $1.15 \times 10^{-4}$ ** |
| Age Range (years) | 20-73 | 22-72 | - | - |
| Gender (Male : Female) | 52 : 18 | 135 : 45 | 0.0136 | 0.9070 |
| UPDRS Off (Mean ± SD) | - | 33.26 ± 8.92 | - | - |
| UPDRS On (Mean ± SD) | - | 17.85 ± 6.27 | - | - |
| H&Y (Mean ± SD) | - | 2.07 ± 1.46 | - | - |
| Age at Onset (Mean ± SD) | - | 48.27 ± 9.14 | - | - |

**significant at 99% confidence interval using t-test

### 2.2. Brain MRI protocol and Image Pre-processing

The structural MRI scans were generated using a 3 Tesla Philips scanner equipped with a 32-channel head coil. These T1-weighted brain MRI scans were obtained using a magnetization prepared rapid acquisition gradient echo sequence (repetition time (TR) of 8.2 ms, echo time (TE) of 3.7 ms, flip angle of 8 degrees, 165 sagittal slices, and a voxel size of $1*1*1mm^3$. The preprocessing has been explained elsewhere (Samantaray et al., 2022a). Thus, the initial images, acquired in DICOM (Digital Imaging and Communications in Medicine) format, were converted into NIfTI (Neuroimaging Informatics Technology Initiative) format using dcm2nii tool (https://www.nitrc.org/projects/dcm2nii/). These were subsequently subjected to preprocessing using Computational Anatomy Toolbox, CAT12 (http://www.neuro.uni-jena.de/cat/), in MATLAB platform (MathWorks, MATLAB 2020a), within the Statistical Parametric Mapping, SPM12 (https://www.fil.ion.ucl.ac.uk/spm/).

Each individual's MRI scans were segmented into distinct tissue types, namely grey matter (GM), white matter (WM), and others. These scans were then normalized to conform to the standard MNI-152 template and modulated through the utilization of the Jacobian determinant. Furthermore, voxel-wise regression was conducted in MATLAB 2021a, using a linear regression analysis to perform, aiming to eliminate the influence of extraneous variables such as age, gender, age and gender interaction. Subsequently, smoothing was done using an 8 mm isotropic Gaussian kernel (Full-width at half maximum) within the SPM environment, separately for GM and WM images.

### 2.3. Subtype identification - using Joint ICA

Patients diagnosed with Parkinson's disease underwent a process of subtyping through the utilization of structural MRI (sMRI) scans. The sequential methodology employed to achieve these subtypes encompassed the following steps:

- The initial stage involved creation of input matrix based on GM and WM volumes.
- This input GM and WM matrix was further decomposed into two separate matrices: a joint loading matrix and a component matrix.
- Components exhibiting significant differences in grey and white matter between Parkinson's patients and healthy controls were then identified.
- From the previously identified components, only those deemed statistically significant were retained.
- PD subtypes were identified by applying a threshold on loading values of the significant components.

The smoothed GM and WM scans were subjected to correlation analysis with the group mean image. This step aimed to identify and eliminate any potential outliers present within the data. The smoothed 3D GM and WM data matrices were separately reshaped to 1D arrays for every individual. Further, these 1D arrays of GM and WM were concatenated horizontally to form a single row vector. This row vector of each individual was concatenated one below the other to form a 2D input matrix. This concatenation process was conducted for all subjects, encompassing 70 healthy controls and 180 Parkinson's disease patients. Joint ICA was implemented using Fusion ICA toolbox (https://trendscenter.org/software/fit/) on the input matrix (250 subject by 1185790 voxel) within MATLAB 2021a platform (MathWorks). Employing an infomax algorithm within the toolbox, spatial independent component analysis (spatial ICA) was applied to decompose the input data matrix (L. Xu, Pearlson, et al., 2009) and number of independent components was fixed to 30 (Zeighami et al., 2015). Thus, it yielded both a joint loading matrix and a component (or source) matrix (Guo et al., 2012b). The loading matrix contains columns that signify the extent to which each subject expresses the joint source differences between the two groups. Within the source matrix, there exist separate matrices for GM sources and WM sources. Every row in the source matrix corresponds to a shared source incorporating both GM and WM maps, displaying identical covariance patterns. These shared sources represent spatial components that are maximally independent.

Furthermore, the loading matrix was employed to identify components displaying significant group differences in grey matter and white matter. Components exhibiting statistically significant intergroup differences were identified using *Student's t-test*. The verification of true positives was carried out by studying the false-discovery rate, (using *fdr_bh* function (Benjamini & Hochberg, 1995) within MATLAB) adhering to a significance level of $P < 0.05$. This utilization of false discovery rate (FDR) controlling procedures assists in curtailing the anticipated proportion of incorrectly rejected null hypotheses, thus mitigating the occurrence of erroneous null hypothesis rejections during multiple testing.

From the identified components, two components with highest effect size were selected (Gupta et al., 2017; Samantaray et al., 2024). The component loadings underwent thresholding using a method adopted from (Rahaman et al., 2020), which demonstrated superior performance compared to other techniques. This thresholding approach involves selecting PD patients having loading coefficients greater than the mean when mean is positive, else selecting PD patients with loadings less than the mean when mean is negative. Thus, PD patients exhibiting higher weight in each of the two components were categorized as two different subtypes, according to the implication that the feature is strongly displayed in those patients (Gupta et al., 2017; Samantaray et al., 2024).

Consequently, PD patients common to both groups were classified as a third subtype. This delineation was aimed at capturing a subtype characterized by shared attributes across both patient and control groups. Thus, the loading matrix played a pivotal role in categorizing subjects into subtypes. On the other hand, the component matrix facilitated the creation of spatial maps corresponding to the components of significance.

### 2.4. Construction of Structural Brain Network

The pre-processed GM and WM scans, used for subtyping, were also considered for construction of subtype-specific structural brain networks and subsequent connectivity assessment (Fig. 1). The network construction involved following steps:

Step 1. **Brain Parcellation:** The first step encompassed the parcellation of GM and WM scans of every PD patients into distinct regions using LONI Probabilistic Brain Atlas, LPBA40 (https://resource.loni.usc.edu/resources/atlases-downloads/) (Shattuck et al., 2008). This process resulted in the division of the maps into 56 GM and 56 WM regions of interest (ROIs) for each PD patient, where each ROI represented a node within the structural network (Step 1 of Fig. 1).

Step 2. **Regional Grey Matter Volume Estimation:** Following brain parcellation, the estimation of regional grey matter volume (rGMV) and regional white matter volume (rWMV) for each of these regions was carried out (Step 2 of Fig. 1). The CAT12 software facilitated in extraction of rGMV and rWMV values for every ROI. Subsequently, separate rGMV and rWMV matrices were generated (with $n * 56$ dimension), where n represents number of subjects within a given subtype.

Step 3. **Association Matrix Construction:** An association matrix was constructed (Step 3 of Fig. 1) by measuring the interregional tissue correlations between different brain regions. For each subtype, a 56 by 56 association matrix was established using Pearson correlation (Samantaray et al., 2022b, 2023; Zalesky et al., 2012) between rGMV and rWMV. In this matrix, each unit ($r_{ij}$) signifies the correlation between GMV of region $i$ and WMV of region $j$, effectively representing an edge within the network.

Step 4. **Adjacency Matrix Construction:** An adjacency matrix was generated (Step 4 of Fig. 1) by binarizing the association matrix using mutual K-nearest neighbor (MKNN) approach (Fig. 2). We propose this approach to be used for thresholding the association matrix based on highest correlation. Here K-nearest neighbor indicates high correlation coefficient. Hence in our MKNN-based thresholding, an edge exists between two nodes ($i$ and $j$), if $i$ is one among the K-nearest neighbor of $j$ and vice versa. To obtain this, primarily, the correlation coefficient of rGMV of every node to rWMV of every other node is sorted in descending order. Further, only the top K number of correlation coefficients are considered between the nodes. Although there are multiple ways to find the K parameter, we have used $K = \sqrt{N}$, where N is the total number of nodes (Hassanat, 2014). We opted for K to be not greater than $\sqrt{N}$, and hence considered K = 7. Hence, an edge exists between nodes $i$ and $j$, if $i$ is amongst the top K correlation coefficients with $j$, and $j$ is also amongst the top K correlation coefficients with $i$. This is explained using an example in Fig. 2. The construction of association and adjacency matrices for the subtypes was accomplished using self-written MATLAB (MathWorks R2021a, https://in.mathworks.com/).

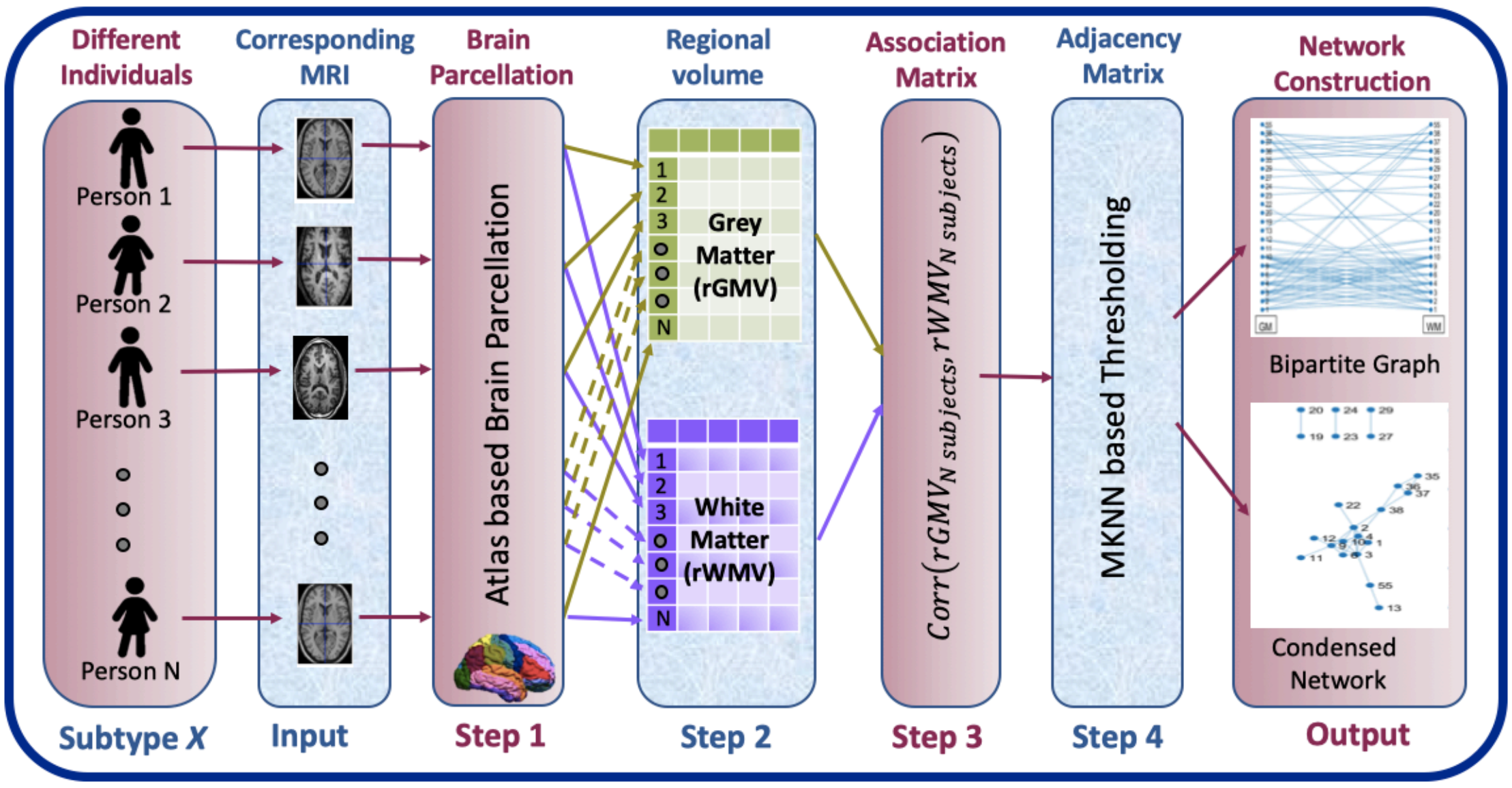

**Figure 1.** Brain network construction using MKNN. This pipeline is applied on each subtype separately. Subtype $X$ is any of the derived subtypes.

Step 5. **Network Construction and Visualization:** The MKNN_NetworkThresholding, an in-house developed tool (https://github.com/NeuralLabIITGuwahati/MKNN_NetworkThresholding), is used to construct and visualize the brain network (Output of Fig. 1). The interconnections between GM and WM nodes are shown as bipartite graphs. The network hubs were visually represented using BrainNet Viewer (Xia et al., 2013).

### 2.5. Network Metrics and Analysis

The adjacency matrix, containing mutual connections, was evaluated to enable comprehensive network analysis. To capture the network's topological characteristics, specific network metrics including average degree, density, betweenness centrality, participation coefficient, clustering coefficient, global efficiency, path length, eigen vector centrality, assortativity, transitivity, gamma and lambda were computed. These network metrics were computed using functions from the Brain Connectivity Toolbox (Rubinov & Sporns, 2010) (https://sites.google.com/site/bctnet/), and explained in Samantaray et al., 2022 (Samantaray et al., 2022b).

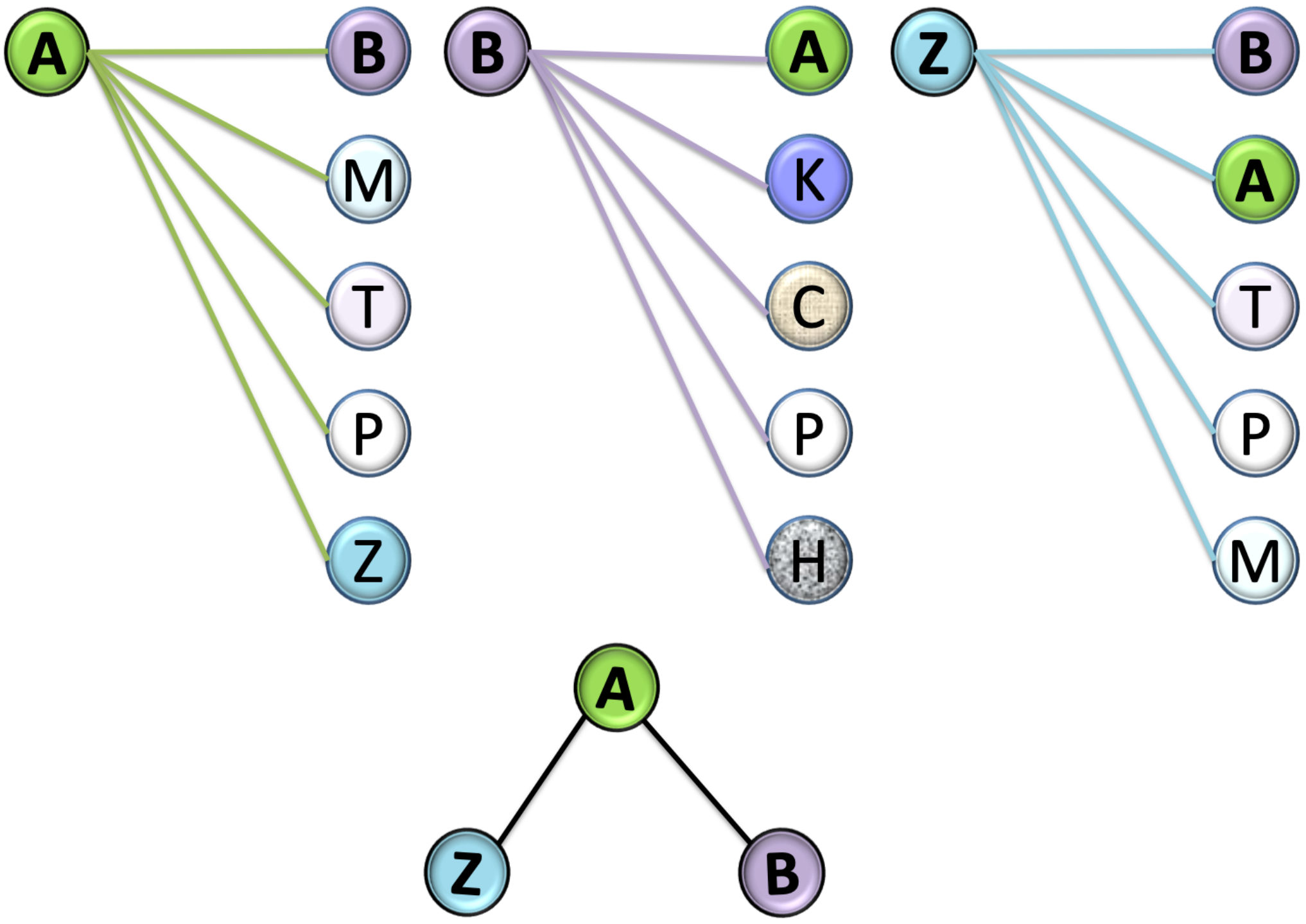

**Figure 2.** Explanation of MKNN-based connectivity. In the association matrix, there exists no zero value, indicating presence of a correlation value between rGMV and rWMV of any pair of ROIs. This implies the network, in such case, is fully connected with edges between every pairs of nodes. To binarize the matrix, the first step is to sort the correlation coefficients in descending order. Considering the nodes indicated in color-coded circles and assuming K = 5 for this figure, we can observe node B is among the five nearest neighbors of node A (Top left), and node A is also among the five nearest neighbors of node B (Top middle), indicating presence of connection between nodes A and B (Bottom figure). Similarly, we can observe node A is among the five nearest neighbors of node Z (Top right), and vice-versa (Top left), indicating A and Z are mutually connected (Bottom figure). On the contrary, although node B is also among the five nearest neighbors of node Z (Top right), node Z is not among the five nearest neighbors of node B (Top middle), indicating absence of connection between nodes B and Z (Bottom figure). Bottom figure shows network with only mutual (bidirectional) connections.

Regions demonstrating degrees and betweenness centrality values exceeding one standard deviation above their respective means were identified as hubs. These findings were generated using MATLAB and subsequently visualized with customized settings in the BrainNet Viewer tool, available at https://www.nitrc.org/projects/bnv.

We also computed the unique brain network identification number (UBNIN) (Samantaray et al., 2023) of individuals belonging to each PD subtypes. As this study considers both GM and WM information, individual brain networks were constructed using correlation ($r_{ij}$) between rGMV for a brain region and rWMV of another. Equation 1 explains this correlation, where $rGMV_i$ is the regional GM volumes of regions $i$ and $rWMV_j$ is the regional WM volumes of regions $j$.

$$r_{ij} = \frac{1}{\left(rGMV_i - rWMV_j\right)^2 + 1} \quad \text{Eqn 1}$$

## 2.6. Statistical Analysis

A *Student's t-test* was conducted to compare differences in age between HC and PD, and Chi Square ($\chi 2$) test for gender, with a significance level of $P < 0.05$. Additionally, *t-test* was also used to assess the differences in joint loading coefficients between HC and PD for every component. The effect size of the components was determined using *Hedge's g test* (Hedges & Olkin, 2014; Lakens, 2013). Notably, *Hedge's g* is recommended when dealing with sample sizes greater than 20, as it offers advantages over the *Cohen's d test*. A bias-corrected version of *Hedge's g* was used to determine the effect size of the differences between PD and HC groups, computed according to the equation 2.

$$Hedge's\ g = \frac{m1 - m2}{s} * \left(1 - \frac{3}{4(n1 + n2) - 9}\right) \quad \text{Eqn 2}$$

where "m1" and "m2" represented mean scores of HC and PD groups, respectively. The values "n1" and "n2" denoted the number of subjects in the HC and PD groups, while "s" stood for the standard deviation across the two samples. These statistical analyses were performed using MATLAB 2021a platform. For comparing the derived subtypes, clinical measures were analyzed. Analysis of variance (ANOVA) was carried out on subtype-specific demographic and clinical variables, including age (in years), Hoehn and Yahr scale (H&Y), Unified Parkinson's Disease Rating Scale at on (UPDRS On) and Off state (UPDRS Off). ANOVA was also applied separately to subtype-specific loading coefficients of significant components. Exploring the associations between these clinical variables and subtype-specific loading coefficients involved using *Pearson* correlation tests. For these correlations to be deemed significant, they needed to exhibit a *P* value below 0.05. The inter-subtype differences in connectivity metrics obtained from the brain network were computed using *Student's t-test*. The MKNN_NetworkThresholding, an in-house developed tool (https://github.com/NeuralLabIITGuwahati/MKNN_NetworkThr-esholding), is used to identify subtypes and assessment of inter-subtype differences in network metrics

## 3. Results

### 3.1. Participants Demographics

Seventy healthy control subjects and 180 Parkinson's patients, ranging in age from 20-73 and 22-72 years, respectively, were examined (Table 1). From a two-sample Student's t-test, it was determined that the age difference between the PD patients and the healthy control participants was statistically significant (t= -3.92, P = $1.15*10^{-4}$). In the HC dataset, there were 52 men and 18 women, while the PD dataset had 135 men and 45 women. The gender chi-squared test revealed no discernible difference between HC and PD ($\chi2$ =0.014, P=0.91).

### 3.2. Joint ICA-based subtyping

The joint loading matrix, obtained upon decomposition, was 250 subjects by 30 components. The loading parameters were analyzed using two-sample t-test for finding differences between PD and HC. Significant difference in loading parameters between HC and PD was identified in twenty three components. Of these, *Hedge's g test* suggested component 23 (*Hedge's g* = 3.5257) had highest effect size, followed by component 28 (*Hedge's g* = 3.5195), Table 2 and Fig. 3.

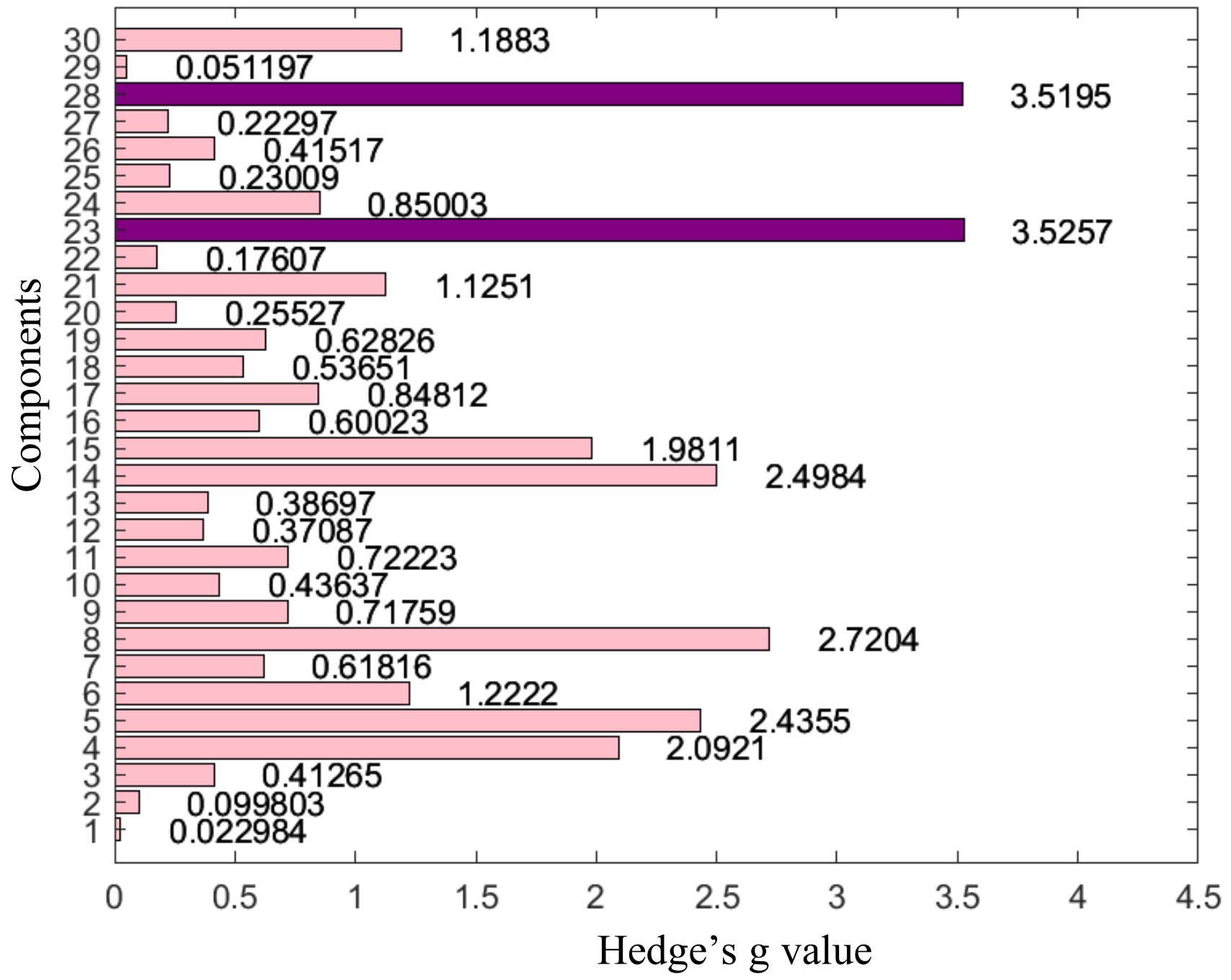

**Figure 3.** Components' effect size for differences between healthy and patient groups

The positive GM map in component 23 covered lentiform nucleus and thalamus, while the negative GM map in the same component included inferior frontal gyrus and rectal gyrus. On the other hand, the positive WM map in component 23 comprised culmen, while negative WM map of the same component involved extranuclear and lentiform nucleus. Similarly, the positive GM map in component 28 covered superior and middle temporal gyri and inferior parietal lobule, while the negative GM map in the same component included subgyral and supramarginal regions. Both positive and negative WM maps in component 28 had the complementary regions as reported in GM maps of the same component, i.e. positive WM map covered subgyral and supramarginal regions, while negative WM map covered superior and middle temporal gyri and inferior parietal lobule. These two component maps are shown in Fig. 4 and the component regions are listed in Table 2.

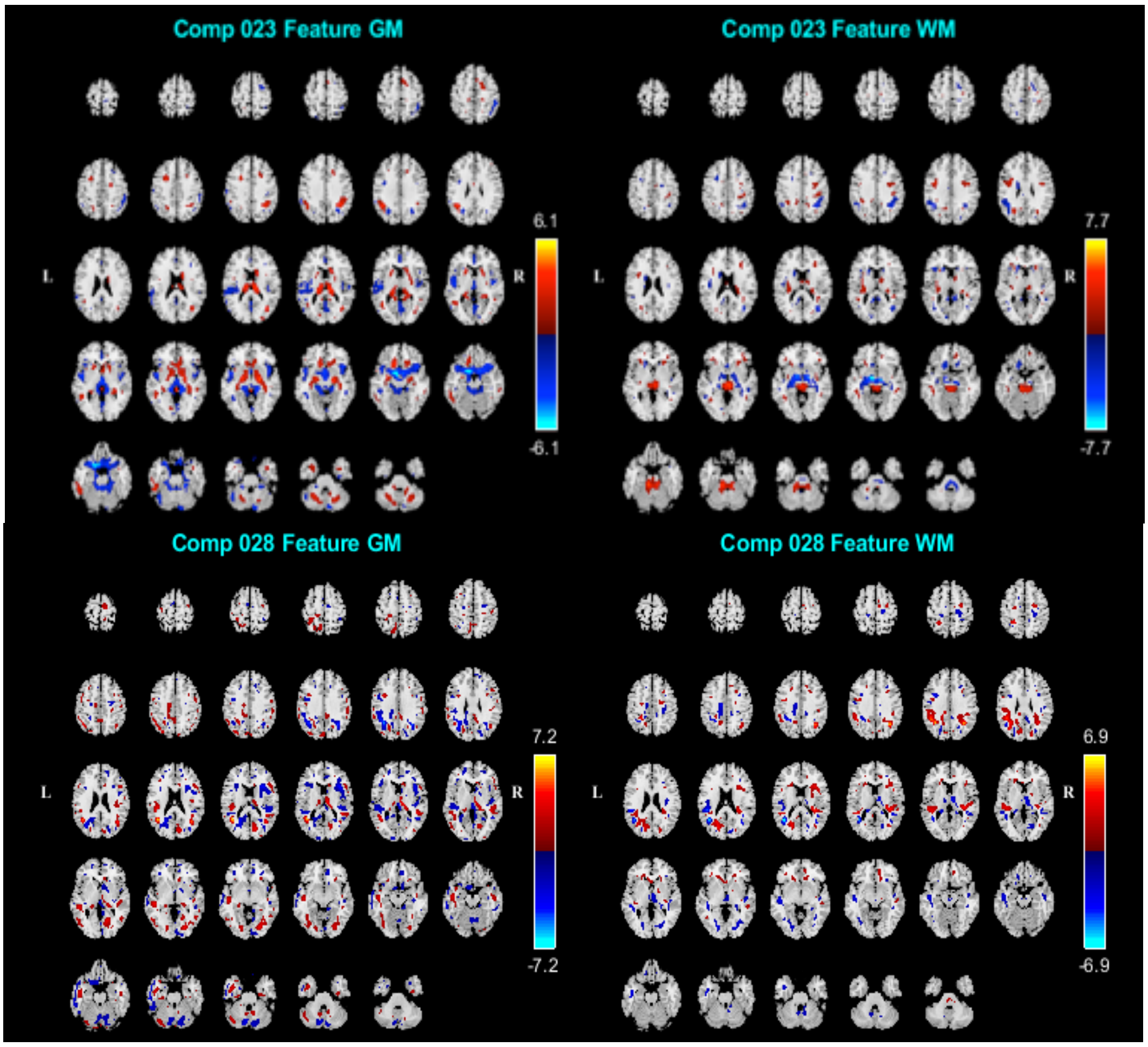

**Figure 4.** Spatial maps of components showing group effect from Fusion ICA toolbox. Voxels above threshold (|Z| > 3.5) are shown. Left panel shows GM regions and right panel shows WM regions. Top and bottom panel shows maps of component 23 and 28, respectively

**Table 2.** Talairach labels of components showing group differences

| GM Talairachs of component 23 using JICA | | | | | |
|---|---|---|---|---|---|
| Talairach | Region (Brodmann Area) | Volume (cm$^3$) L/R | Max value (x,y,z) L/R | MNI (x,y,z) L/R | *Hedge's g* effect size |
| Positive | Lentiform Nucleus | 0.1/ 0.3 | 3.7 (-20,-8, -6)/ 4.4 (20, -10, -6) | (-20,-8,-8)/ (20, -10, -8) | 3.5257 |
| | Thalamus | 0.2/ 0.4 | 3.9 (-12, -19, 14)/ 4.3 (12, -23, 14) | (-12, -20, 14)/ (12, -24, 14) | |
| Negative | Inferior Frontal Gyrus (47) | 0.3/ 0.5 | 5.2 (-22, 12, -22)/ 4.3 (24, 15, -19) | (-22, 14, -26)/ (24, 16, -22) | |
| | L Rectal Gyrus (11) | 0.4/ 0.0 | 5.1 (-10, 12, -22) | (-10, 14, -26) | |
| **WM Talairachs of component 23 using JICA** | | | | | |
| Positive | Culmen | 0.3/ 0.3 | 5.4 (-12,-32,-20)/ 4.6 (14, -34, -20) | (-12, -32, -26)/ (14, -34, -26) | |
| Negative | Extra-Nuclear | 1.0/ 0.5 | 6.1 (-12, -4, -8)/ 5.6 (12, -4, -8) | (-12, -4, -10)/ (12, -4, -10) | |
| | R Lentiform Nucleus | 0.0/ 0.3 | 5.2 (18, -8, -6) | (18, -8, -8) | |
| **GM Talairachs of component 28 using JICA** | | | | | |
| Positive | Superior Temporal Gyrus (22,39) | 1.2/0.4 | 6.7 (-40,-54,17)/ 4.1 (42, -55, 21) | (-40,-56,16)/ (42, -58, 20) | 3.5195 |
| | Middle Temporal Gyrus (19,22,39) | 0.5/0.3 | 5.6(-38,-57,18)/ 5.2 (42, -59, 21) | (-38, -60, 16) /(42, -62, 20) | |
| | L Inferior Parietal Lobule (7, 40) | 0.4/0.0 | 4.4(-42,-56,47) | (-42,-60,48) | |
| Negative | Sub-Gyral | 0.6/ 0.4 | 5.7(-34,-53,32)/ 5.1 (36, -55, 32) | (-34,-56,32)/ (36, -58, 32) | |
| | L Supramarginal | 0.5/ 0.0 | 4.4(-42,-43,30) | (-42,-46,30) | |
| **WM Talairachs of component 28 using JICA** | | | | | |
| Positive | Sub-Gyral (22,39) | 0.6/0.4 | 5.8(-34,-53,32)/ 5.3 (36, -55, 32) | (-34,-56,32)/ (36, -58, 32) | |
| | Supramarginal | 0.6/0.1 | 4.9(-44,-43,30)/ 4.0 (40, -53, 34) | (-44,-46,30)/ (40, -56, 34) | |
| Negative | Superior Temporal Gyrus (22,39) | 1.0/0.1 | 6.1(-40,-54,17)/ 4.0 (42, -55, 21) | (-40,-56,16)/ (42, -58, 20) | |
| | Middle Temporal Gyrus (19,39) | 0.6/0.1 | 5.6(-38,-57,18)/ 4.7 (42, -59, 20) | (-38,-60,16)/ (42, -62, 18) | |
| | L Inferior Parietal Lobule | 0.2/0.0 | 4.7 (-48,-34,22) | (-48, -36, 22) | |

L: Left; R: Right, MNI: Montreal Neurological Institute standard atlas

The joint loadings of significant components were thresholded as per N-way biclustering algorithm (Rahaman et al., 2020) to derive subtypes within PD. The following subtypes were obtained:

1) Subtype A (51 PD patients, with joint loading coefficients highly expressed in component 23),
2) Subtype B (57 PD patients, with joint loading coefficients highly expressed in component 28), and
3) Subtype AB (36 PD patients, with joint loading coefficients highly expressed in both component 23 and 28).

The loading distribution for healthy subjects and various PD subtypes before and after subtyping in component 23 are plotted (Fig. 5) against those of component 28.

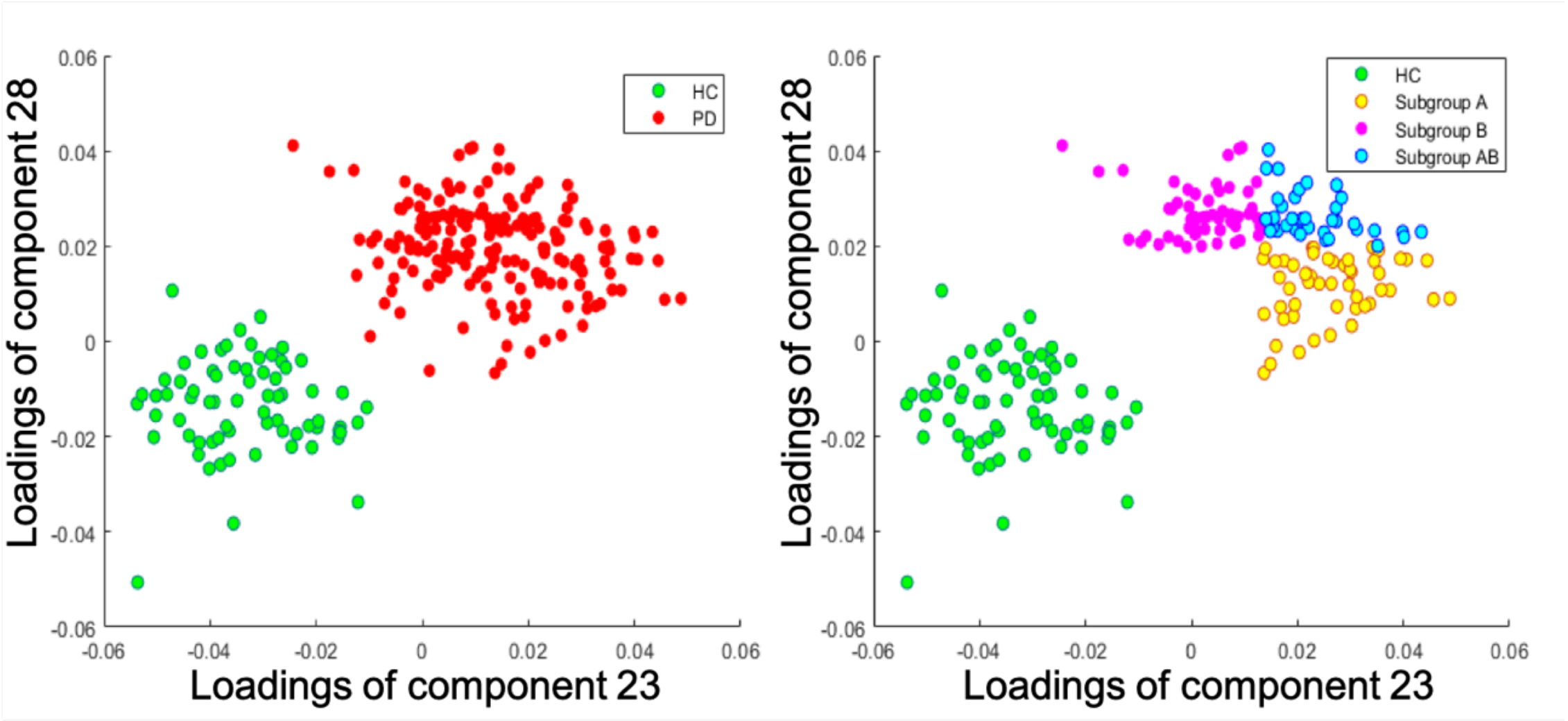

**Figure 5.** Plot of component Loadings of HC and various PD subtypes before (left panel) and after (right panel) subtyping.

### 3.2.1. Loading Differences

The joint loadings between three subtypes were significantly different in both component 23 ($P$ = 3.22 * $10^{-33}$) and component 28 ($P$ = 1.32 * $10^{-30}$) at P < 0.01 (Fig. 6). Table 3 summarizes the demographic and clinical data of every PD subtype. A significant difference was observed in UPDRS Off ($P$ = 0.0076), UPDRS On ($P$ = 0.0182), H&Y ($P$ = 0.0384), age ($P$ =3.3* $10^{-5}$) and age at onset ($P$ = 0.0004) between the subtypes at 95% confidence interval (Fig. 6).

#### 3.2.2. Volume Differences

No difference was observed in total GM volume or total WM volume in any pairs of subtypes. The inter-subtype difference in total GM volume was found to be non-significant using *Student's t-test*, i.e. between subtypes A and B ($t$ = 1.12, $P$ = 0.27), A and AB ($t$ = -0.06, $P$ = 0.95), B and AB ($t$ = -1.06, $P$ = 0.29). Similarly the total WM volume difference was also insignificant between subtypes A and B ($t$ = 0.59, $P$ = 0.55), A and AB ($t$ = -0.98, $P$ = 0.33), B and AB ($t$ = -1.37, $P$ = 0.17).

#### 3.2.3. Clinical Differences

The loading coefficients of component 23 was observed to be correlated with UPDRS Off ($r$ = 0.3, $P$ = 0.03), age at onset ($r$ = 0.27, $P$ = 0.06), age ($r$ = 0.36, $P$ = 0.009) of subtype A-specific subjects. Similarly, UPDRS Off ($r$ = 0.26, $P$ = 0.05), H&Y ($r$ = 0.24, $P$ = 0.07) and age ($r$ = 0.29, $P$ = 0.03) of subtype B-specific subjects were found to have association with loading coefficients of component 23, and age ($r$ = 0.22, $P$ = 0.10) with loading coefficients of component 28. However, none of the clinical variables of subtype AB-specific subjects had significant association with loading coefficients of either component 23 or of component 28.

### 3.3. Brain network of PD subtypes

The association matrix for every subtype comprised of 56 rows and 56 columns corresponding to 56 brain regions, i.e. nodes. Each cell contained the correlation between regional GM and WM volumes. Using the MKNN_NetworkThresholding, a binary-undirected graph is constructed for subtype A, B and AB. This resulted in 34 edge network, 40 edge network and 33 edge network respectively. These result to bipartite graphs explaining connections between GM and WM nodes. Each of the subtypes has a few disconnected nodes when mutual connections are only taken into account. Further, these completely disconnected nodes are removed from the respective adjacency matrices and bipartite graphs for better visualization (Fig. 7). Thereby, 22 nodes were interconnected in subtype A, 25 nodes in subtype B and 24 nodes in subtype AB due to mutual connections.

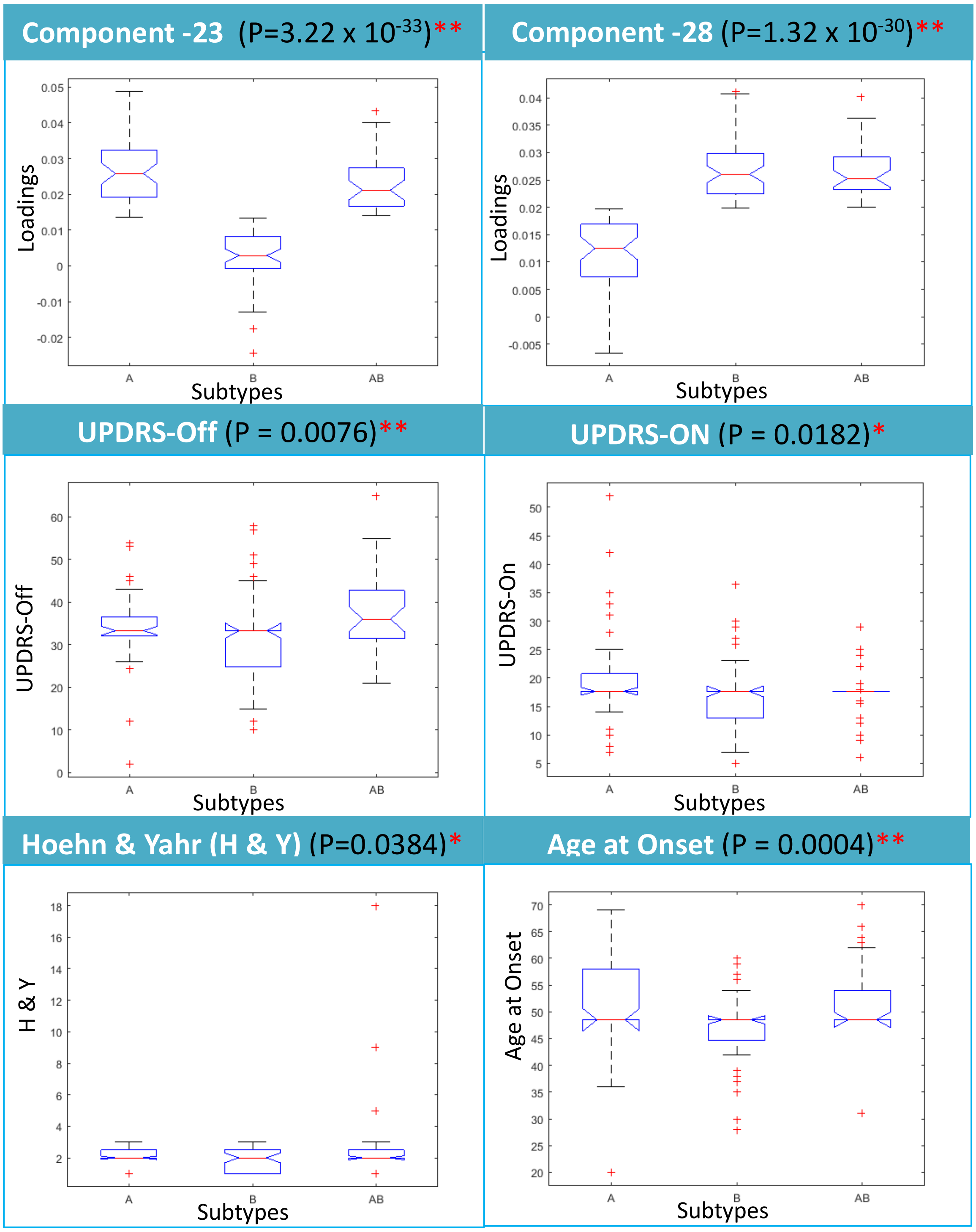

**Figure 6.** ANOVA of subtype-specific loading coefficients and clinical variables
* Significant at 95% confidence interval; ** Significant at 99% confidence interval

**Table 3.** Subject demographics for JICA based subtypes

| Subtypes | A exclusively | B exclusively | AB | p-value |
|---|---|---|---|---|
| Count | 51 | 57 | 36 | - |
| Gender (F:M) | 18:33 | 9:48 | 5:31 | - |
| Age | 59.0 ± 8.38 | 52.3 ± 8.66 | 59.3 ± 8.35 | $3.33 * 10^{-5}$** |
| Age Range | 32-72 | 32-70 | 37-72 | - |
| UPDRS Off | 33.94 ± 8.05 | 31.01 ± 10.2 | 37.29 ± 9.53 | 0.0076** |
| UPDRS On | 19.92 ± 7.59 | 16.65 ± 5.9 | 17.11 ± 4.27 | 0.0182* |
| H & Y | 2.03 ± 0.6 | 1.9 ± 0.67 | 2.7 ± 2.9 | 0.0384* |
| Age at Onset | 52.0 ± 8.35 | 46.63 ± 6.89 | 51.65 ± 7.09 | 0.0004* |
| Total GMV ($cm^3$) | 511.04 ± 45.82 | 500.22 ± 53.76 | 511.59 ± 44.22 | 0.4145 |
| Total WMV ($cm^3$) | 315.82 ± 33.41 | 311.32 ± 43.37 | 323.00 ± 33.76 | 0.3502 |

Age, UPDRS, H&Y, Total GMV (grey matter volume) and Total WMV (white matter volume) are expressed in mean ± SD, * significance at 95% Confidence interval, ** significance at 99% Confidence interval.

The density of binary network was found to be 0.0221 in subtype A, 0.0260 in subtype B and 0.0214 in subtype AB (Fig. 8). The average network degree was observed to be 1.2143 for subtype A, 1.4286 for subtype B and 1.1786 for subtype AB (Fig. 8). The global efficiency of subtype A was 0.0439, subtype B was 0.0821 and subtype AB was 0.0398 (Fig. 8). The average assortativity coefficient for subtype A (assortativity = 0.3513) was slightly higher than subtype AB (assortativity = 0.3506) for subtype AB and that for subtype B (assortativity = 0.2352) was observed to be the least (Fig. 8). A similar trend was notices for transitivity, where the coefficient was found to be 0.5909 for subtype A, 0.5042 for subtype AB and 0.3488 for subtype B (Fig. 8).

*Student's t-test* suggests significant group difference in betweenness centrality between subtypes A and B ($t$ = -2.34, $P$ = 0.02) and between subtypes B and AB ($t$ = 0.006, $P$ = 2.79). Participation coefficient significantly differed between subtypes A and B ($t$ = -2.16, $P$ = 0.03). However, no significant difference was observed in clustering coefficient, gamma, or eigen vector centrality between the subtype-specific networks.

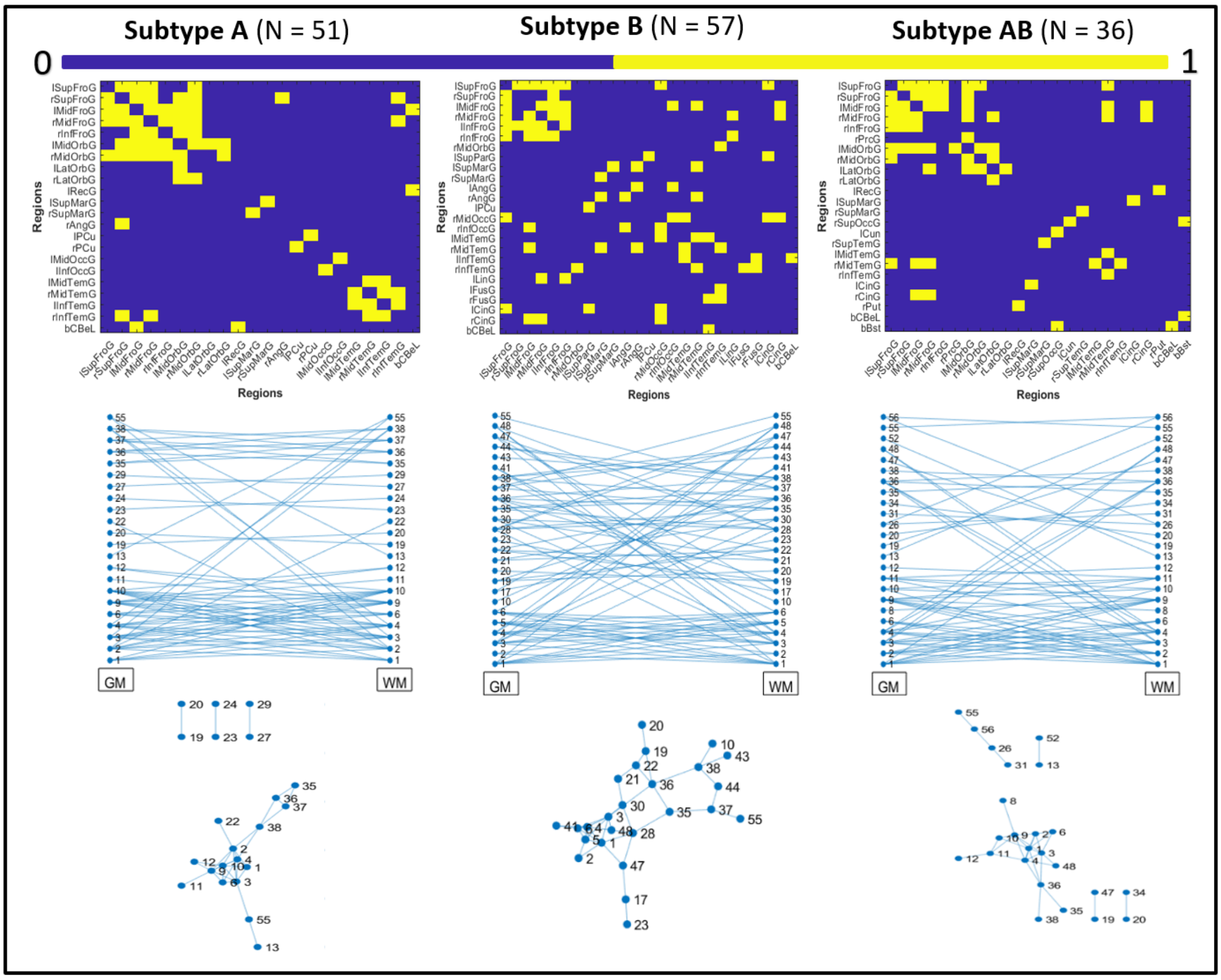

**Figure 7.** Subtype- specific mutual k-nearest neighbor-based adjacency Matrices, bipartite graph and condense network. Adjacency matrices are shown in top row. Middle row shows bipartite graphs indicating mutual connections between GM regions (left nodes indicated by GM) and white matter regions (right nodes indicated by WM). Bottom row shows condensed network, where the same grey matter region (eg. node no. 13 of bipartite) and white matter region (eg. Node no. 13 of bipartite) are indicated by a single region (node no. 13) in condensed network. The node numbers are indicate regions, mentioned in the appendix.

Hubs were deciphered from betweenness centrality and average degree of the networks (Fig. 9). The hubs identified are right superior frontal gyrus (rSFG, royal blue color) in subtype A, right middle occipital gyrus (rMOG, magenta color) in subtype B and left lateral orbitofrontal gyrus (lLatOrbG, green color) in subtype AB. A few common hubs were also found in different subtypes, such as right inferior temporal gyrus (rITG, cyan color) in Subtypes A and B, left caudate (lCaud, mustard yellow color) in subtypes A and AB, left superior frontal gyrus (lSFG, dark red color) and right middle temporal gyrus (rMTG, dark red color) in subtypes B and AB. Additionally, left middle frontal gyrus (lMFG, grey color) was found in all three subtypes, which suggests this conserved brain region might be keeping communication active in all Parkinson's patients.

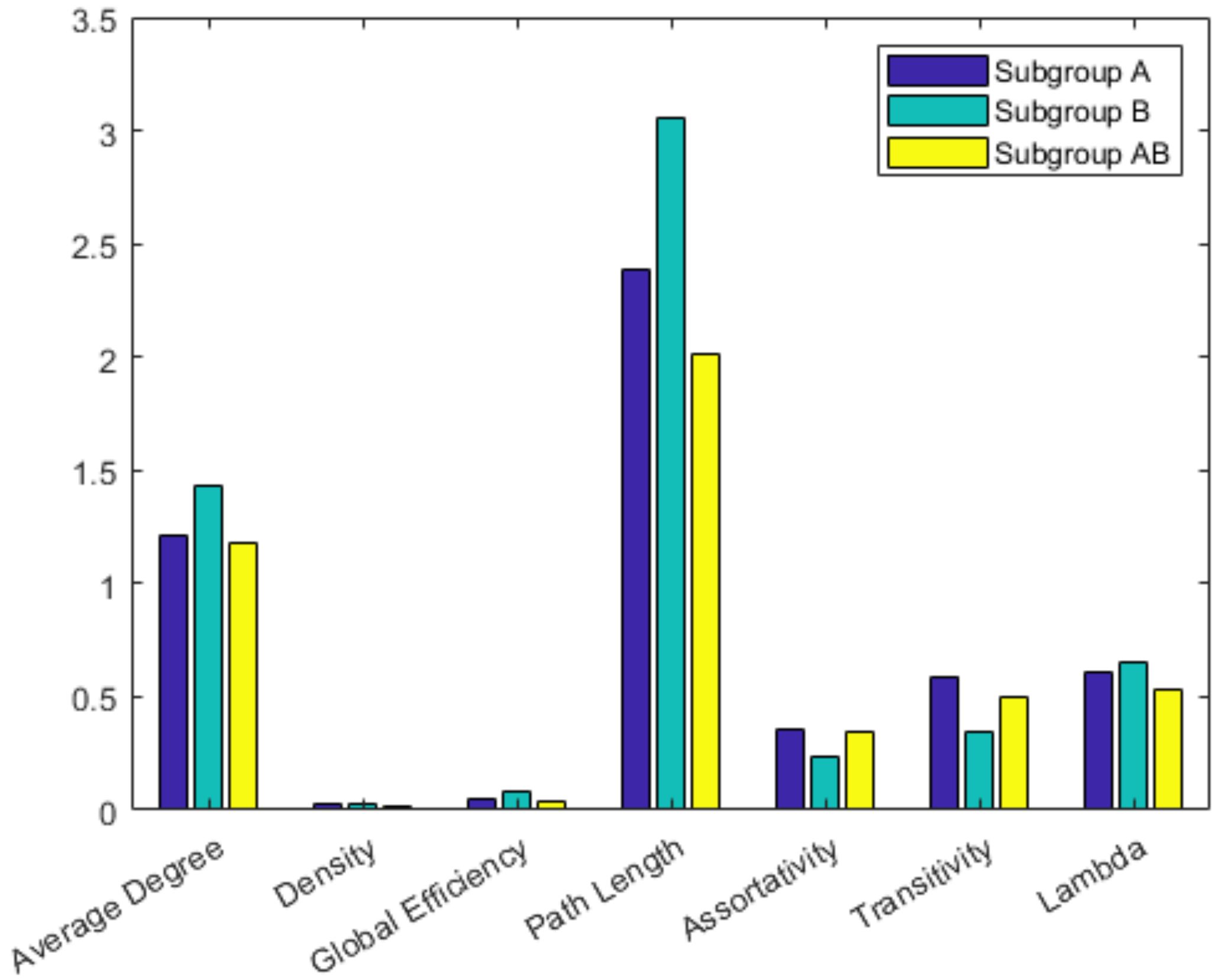

**Figure 8.** Plot showing network metrics of JICA-based subtypes

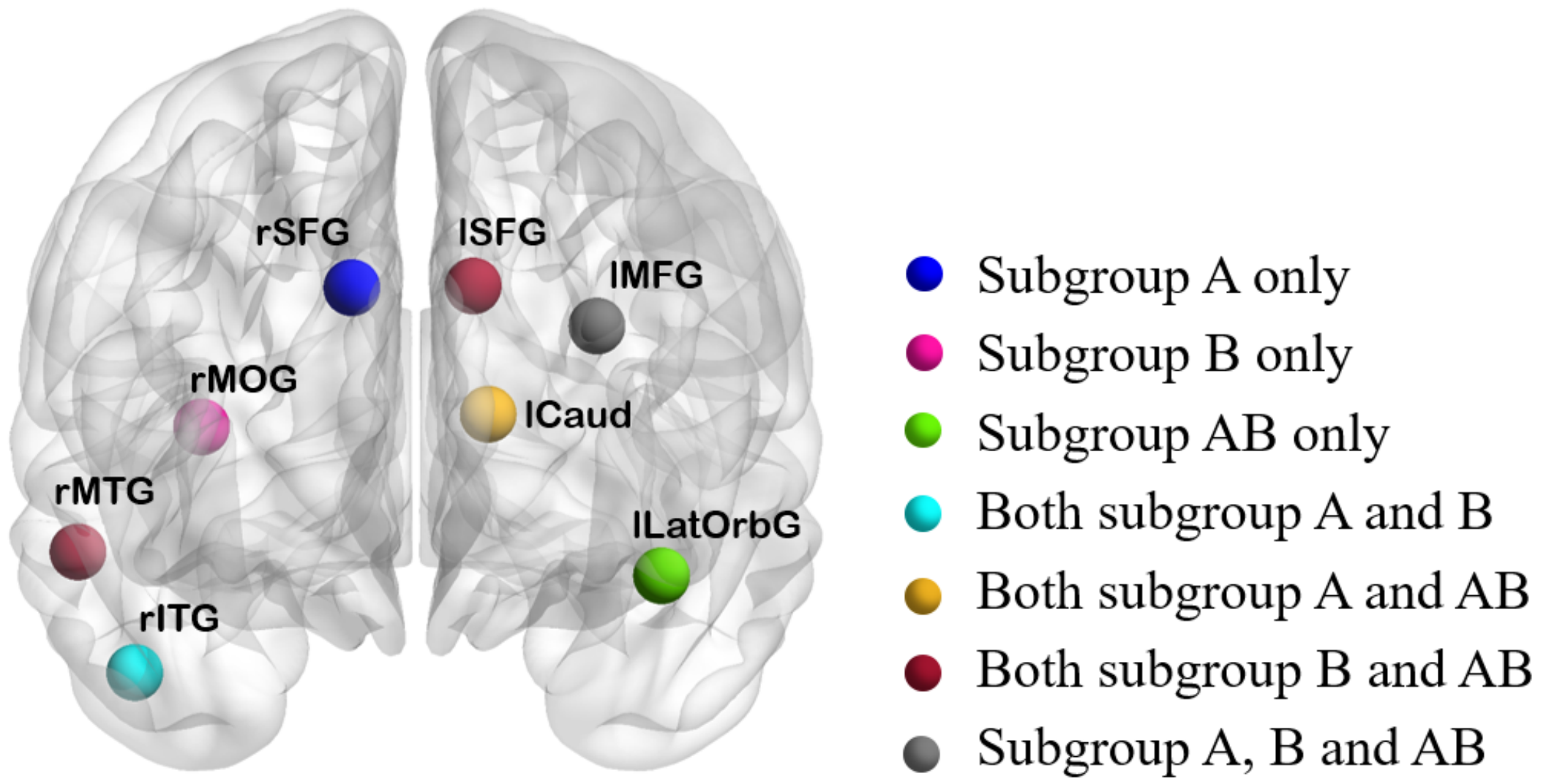

**Figure 9.** Hub regions in different subtypes

Individuals of each subtype (Subtype A: N = 51; Subtype B: N = 57; and Subtype AB: N = 36) were observed with distinct UBNIN values. The violin plot (Fig. 10) displays the distribution of UBNIN for subtype-specific PD individuals.

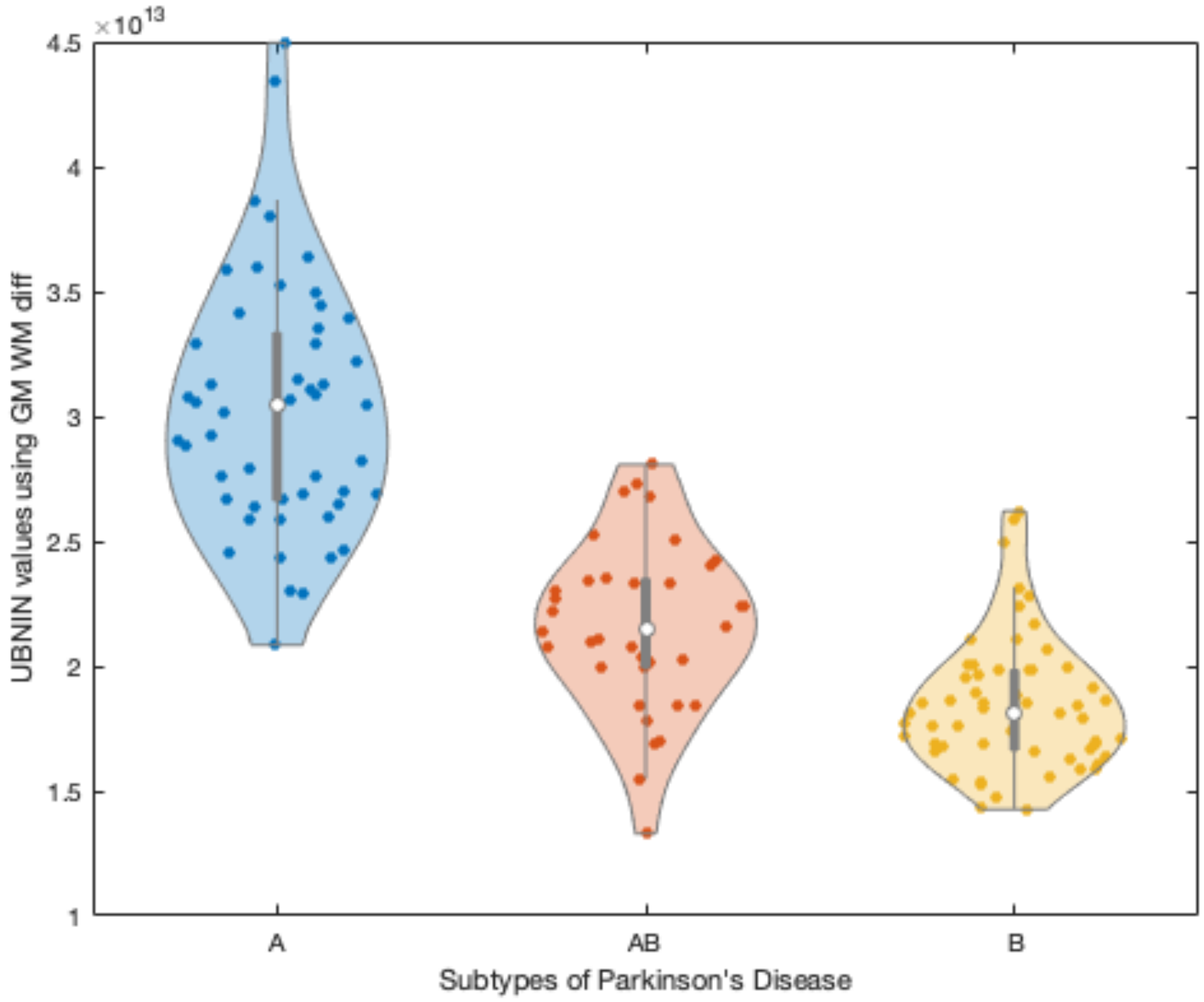

**Figure 10.** Violin plot demonstrating UBNIN distribution of subtype-specific PD individuals

## 4. Discussions

Although other multimodal methods are available for different applications (Calhoun & Sui, 2016; Luo et al., 2024), joint ICA is a simple method enabling data fusion of intra-modality information. We performed a joint ICA approach to identify joint GM and WM components differing significantly between healthy controls and PD patients. Considering the components with top two effect sizes, the PD patients were categorized based on the weights in the respective components.

- The first PD category, i.e., subtype A, weighed higher in the component with highest effect size, which was component 23.
- The second PD category, i.e., subtype B, weighed higher in the component with second highest effect size, which was component 28.

- Patients belonging to the third PD category, i.e., subtype AB, weighed higher in both the components, i.e., component 23 and 28.

The above subtypes were obtained from fused information of grey and white matter tissues. It's intriguing to notice that while assessing these subtypes with the clinical features, we observed that patients belonging to:

- subtype B had significantly least chronological age, UPDRS-Off value, UPDRS-On value, H&Y scale and age at onset than the other subtypes B and AB.
- subtype A had chronological age, UPDRS-Off value, UPDRS-On value and H&Y scale higher than those of subtype B, but lesser than those of subtype AB.
- subtype AB had significantly highest chronological age, UPDRS-Off value, UPDRS-On value and H&Y scale than the other subtypes A and B.

UPDRS-Off assesses the severity of motor symptoms at no medication state to remove the effect of dopaminergic medications . UPDRS-On measures the motor symptoms of the patients after drug was given. H&Y scale, the staging system indicating PD severity (Hoehn & Yahr, 1967), is a standard for measuring disability and impairment (Goetz et al., 2004).

Hence, our study results suggest that subtype B is the mild-severe type with mild motor symptoms, subtype A is the intermediate type, subtype AB is the most-severe type with predominance in motor impairment. Age at onset shows the age at which motor symptoms appeared.

The joint loading coefficients were also correlated with clinical features of the subtypes to understand the association between them. The investigation suggested loading coefficient of joint component 23 in subtype A specific subjects were positively correlated with UPDRS-Off, age at onset and chronological age. Similarly, subtype B subjects' loading coefficients of joint component 23 were positively correlated with UPDRS-Off, H&Y and chronological age. These subjects' loading coefficients in joint component 28 were also positively correlated with age.

The brain network of each of these derived subtypes were constructed based on association between grey matter and white matter volume of every pairs of brain regions (nodes). Each of the association matrices, hence obtained, were binarized using mutual K-nearest neighbor (MKNN) based threshold. MKNN is a node affinity measure that defines a bidirectional relationship between the nodes based on similarity values between grey matter and white matter information. Network efficiency indicates the number of steps needed for information transfer from one brain region to another. Highly efficient networks are thought to take fewer steps or shorter path lengths than inefficient networks, which usually take several steps. The global efficiencies show that subtype B transmits information faster than subtype A, and subtype A faster than AB. Also, path length values of subtype B is greater than that of subtype A, which is again higher than that of subtype AB. So, global efficiency and pathlength results indicate that subtype B might transfer information efficiently through longer path.

A measure of a network's resilience, network assortativity is the association between the degrees of all nodes on two opposite sides of an edge (Rubinov & Sporns, 2010). In assortative networks, vertices with higher degrees tend to be connected together in a sub-network with greater mean degree than the overall network (Newman, 2002). This phenomenon is reflected in the network structure of the three subtypes (Fig. 7) where subtype A with highest assortativity has more specific subnetwork comprising less number of nodes as compared to that of subtypes AB and B. Corroborating with the phenomenon, same nodes on opposite hemispheres are interconnected, *eg*. In subtype A, Left and right supramarginal gyrus (nodes 19 and 20 in Fig. 7, bottom left), left and right precuneus (nodes 23 and 24 in Fig. 7, bottom left), left and right middle temporal gyrus (node 35 and 36 in Fig. 7, bottom left), left and right inferior temporal gyrus (nodes 37 and 38 in Fig. 7, bottom left), left and right superior frontal gyrus (nodes 1 and 2 in Fig. 7, bottom left), left and right middle frontal gyrus (nodes 3 and 4 in Fig. 7, bottom left) and left and right middle orbitofrontal gyrus (nodes 9 and 10 in Fig. 7, bottom left). High-degree and low-degree nodes in assortative networks prefer connecting to other high- and low-degree nodes, respectively (Newman, 2002). This order tends to erode as the assortativity coefficient falls, and some vertices start forming new connections with vertices that have degrees that are less similar to their own degrees. In PD subtype B, the lower assortativity coefficient might indicate that few regions start establishing or increasing connections with other regions. The decreased assortativity in PD subtype B might suggest that few brain regions begin to connect with other regions, possibly the reason for a bigger subnetwork than those of subtypes A and AB.

Betweenness centrality quantifies involvement of each node in information flow within the network. It also helps examining how different nodes (brain regions) interact with each other. Hence, in order to determine information flow both within and between subnetworks, we employed betweenness centrality as a metric for analysis. Our results indicate subtype B differing in betweenness centrality from the other two subtypes (subtypes A and AB). We also used this measure to find hub regions in the networks. Participation coefficient is another metric which was compared between the subtypes. Participation coefficient distinguishes a provincial hub from a connector hub, and therefore indicates whether a node transmits information to distant nodes or to its neighboring nodes. Our study suggests a significant difference in participation coefficient between subtypes A and B. Frontal and temporal gyrus were majorly found as hubs in different subtypes, showing these regions with significant role in information transmission through the network.

## Challenges and opportunities

There are a few limitations associated with our study. Only two joint sources with highest effect size between healthy and PD were utilized for subtyping, possibly why all PD patients could not be categorized into any of the derived subtypes. Hence, consideration of all components (Shreeram et al., 2023), exhibiting group differences, might help to subtype the entire PD cohort. We used multiple features (GM and WM information) from a single modality. However, multimodal neuroimaging data might be advantageous in PD subtyping. In this direction, imaging transcriptomics, *i.e.* combined evaluation of neuroimaging and gene expression data to understand molecular changes in PD, may offer deep insights into the disease pathology and unravel disease heterogeneity. Advanced approaches on joint ICA (Agcaoglu et al., 2022; Khalilullah et al., 2023) such as parallel multilink JICA and joint non-linear ICA may also be implemented for further studies. Brain network using mutual neighborhood is dependent on K in our study. This limits the average network degree to K. However, the strength of our study is the implementation of a single threshold value instead of a range of thresholds (*eg*. Sparsity, density) based on mutual information between GM (neuronal head) and WM (axon) tissue. Further, more advanced methods such as gravitational mutual nearest neighbor (Ma et al., 2023), adaptive mutual K-NN method (Wang et al., 2023) may be explored in brain connectivity analysis.

## 5. Conclusion

In this study, based on GM and WM information, we successfully deciphered PD subtypes associated with specific clinical symptoms. We also found inter-subtype differences in network metrics based on mutuality between the brain regions. Thus, our approach of identifying subtypes within the disease may be useful in precision treatment in contrast to traditional one-size-fits-all approach, where a single treatment is uniformly administered to all PD patients. Our study encourages tailoring of subtype specific treatments by considering the respective spectrum of clinical symptoms along with neuroimaging features.

### Code availability

The code developed for this work is available at the below GitHub link:
https://github.com/NeuralLabIITGuwahati/MKNN_NetworkThresholding

### Author Contributions

Mutual K-NN thresholding was proposed by VVS to TS and multimodal fusion by CNG to TS. Entire data analysis and programming was done by TS. Data was provided by PKP and JS. All authors were involved in writing and revisions of manuscripts. This work was part of PhD research done by TS at IIT Guwahati .

### Conflict of interest

The authors declare no conflict of interest.

## References

Agcaoglu, O., Silva, R. F., & Calhoun, V. (2022). Multimodal fusion of brain imaging data with joint non-linear independent component analysis. *2022 IEEE 14th Image, Video, and Multidimensional Signal Processing Workshop (IVMSP)*, 1–5. https://doi.org/10.1109/IVMSP54334.2022.9816248

Albrecht, F., Poulakis, K., Freidle, M., Johansson, H., Ekman, U., Volpe, G., Westman, E., Pereira, J. B., & Franzén, E. (2022). Unraveling Parkinson's disease heterogeneity using subtypes based on multimodal data. *Parkinsonism & Related Disorders*, *102*, 19–29. https://doi.org/10.1016/j.parkreldis.2022.07.014

Bassett, D. S., & Sporns, O. (2017). Network neuroscience. *Nature Neuroscience*, *20*(3), 353–364. https://doi.org/10.1038/nn.4502

Benjamini, Y., & Hochberg, Y. (1995). Controlling the False Discovery Rate: A Practical and Powerful Approach to Multiple Testing. *Journal of the Royal Statistical Society. Series B (Methodological)*, *57*(1), 289–300. https://www.jstor.org/stable/2346101

Berg, D., Postuma, R. B., Bloem, B., Chan, P., Dubois, B., Gasser, T., Goetz, C. G., Halliday, G. M., Hardy, J., Lang, A. E., Litvan, I., Marek, K., Obeso, J., Oertel, W., Olanow, C. W., Poewe, W., Stern, M., & Deuschl, G. (2014). Time to redefine PD? Introductory statement of the MDS Task Force on the definition of Parkinson's disease. *Movement Disorders: Official Journal of the Movement Disorder Society*, *29*(4), 454–462. https://doi.org/10.1002/mds.25844

Calhoun, V. D., & Sui, J. (2016). Multimodal fusion of brain imaging data: A key to finding the missing link(s) in complex mental illness. *Biological Psychiatry. Cognitive Neuroscience and Neuroimaging*, *1*(3), 230–244. https://doi.org/10.1016/j.bpsc.2015.12.005

Cao, K., Pang, H., Yu, H., Li, Y., Guo, M., Liu, Y., & Fan, G. (2022). Identifying and validating subtypes of Parkinson's disease based on multimodal MRI data via hierarchical clustering analysis. *Frontiers in Human Neuroscience*, *16*. https://www.frontiersin.org/articles/10.3389/fnhum.2022.919081

Eglit, G. M. L., Lopez, F., Schiehser, D. M., Pirogovsky-Turk, E., Litvan, I., Lessig, S., & Filoteo, J. V. (2021). Delineation of Apathy Subgroups in Parkinson's Disease: Differences in Clinical Presentation, Functional Ability, Health-related Quality of Life, and Caregiver Burden. *Movement Disorders Clinical Practice*, *8*(1), 92–99. https://doi.org/10.1002/mdc3.13127

Fereshtehnejad, S.-M., Zeighami, Y., Dagher, A., & Postuma, R. B. (2017). Clinical criteria for subtyping Parkinson's disease: Biomarkers and longitudinal progression. *Brain: A Journal of Neurology*, *140*(7), 1959–1976. https://doi.org/10.1093/brain/awx118

Goetz, C. G., Poewe, W., Rascol, O., Sampaio, C., Stebbins, G. T., Counsell, C., Giladi, N., Holloway, R. G., Moore, C. G., Wenning, G. K., Yahr, M. D., & Seidl, L. (2004). Movement Disorder Society Task Force report on the Hoehn and Yahr staging scale: Status and recommendations The Movement Disorder Society Task Force on rating scales for Parkinson's disease. *Movement Disorders*, *19*(9), 1020–1028. https://doi.org/10.1002/mds.20213

Guo, X., Han, Y., Chen, K., Wang, Y., & Yao, L. (2012a). Mapping joint grey and white matter reductions in Alzheimer's disease using joint independent component analysis. *Neuroscience Letters*, *531*(2), 136–141. https://doi.org/10.1016/j.neulet.2012.10.038

Guo, X., Han, Y., Chen, K., Wang, Y., & Yao, L. (2012b). Mapping joint grey and white matter reductions in Alzheimer's disease using joint independent component analysis. *Neuroscience Letters*, *531*(2), 10.1016/j.neulet.2012.10.038. https://doi.org/10.1016/j.neulet.2012.10.038

Gupta, C. N., Castro, E., Rachkonda, S., van Erp, T. G. M., Potkin, S., Ford, J. M., Mathalon, D., Lee, H. J., Mueller, B. A., Greve, D. N., Andreassen, O. A., Agartz, I., Mayer, A. R., Stephen, J., Jung, R. E., Bustillo, J., Calhoun, V. D., & Turner, J. A. (2017). Biclustered Independent Component Analysis for Complex Biomarker and Subtype Identification from Structural Magnetic Resonance Images in Schizophrenia. *Frontiers in Psychiatry*, *8*. https://doi.org/10.3389/fpsyt.2017.00179

Hassanat, A. (2014). *Solving the Problem of the K Parameter in the KNN Classifier Using an Ensemble Learning Approach. 12*(8). arXiv preprint arXiv:1409.0919. arXiv preprint arXiv:1409.0919

Hedges, L. V., & Olkin, I. (2014). *Statistical Methods for Meta-Analysis*. Academic Press.

Hoehn, M. M., & Yahr, M. D. (1967). Parkinsonism: Onset, progression and mortality. *Neurology*, *17*(5), 427–442. https://doi.org/10.1212/wnl.17.5.427

Hu, Z., & Bhatnagar, R. (2012). Clustering algorithm based on mutual K-nearest neighbor relationships. *Statistical Analysis and Data Mining*, *5*(2), 100–113. https://doi.org/10.1002/sam.10149

Huppertz, H.-J., Möller, L., Südmeyer, M., Hilker, R., Hattingen, E., Egger, K., Amtage, F., Respondek, G., Stamelou, M., Schnitzler, A., Pinkhardt, E. H., Oertel, W. H., Knake, S., Kassubek, J., & Höglinger, G. U. (2016). Differentiation of neurodegenerative parkinsonian syndromes by volumetric magnetic resonance imaging analysis and support vector machine classification. *Movement Disorders*, *31*(10), 1506–1517. https://doi.org/10.1002/mds.26715

Jarvis, R. A., & Patrick, E. A. (1973). Clustering Using a Similarity Measure Based on Shared Near Neighbors. *IEEE Transactions on Computers*, *C–22*(11), 1025–1034. https://doi.org/10.1109/T-C.1973.223640

Khalilullah, K. M. I., Agcaoglu, O., Sui, J., Adali, T., Duda, M., & Calhoun, V. D. (2023). Multimodal fusion of multiple rest fMRI networks and MRI gray matter via parallel multilink joint ICA reveals highly significant function/structure coupling in Alzheimer's disease. *Human Brain Mapping*, *44*(15), 5167–5179. https://doi.org/10.1002/hbm.26456

Khanna, K., Gambhir, S., & Gambhir, M. (2023). A novel technique for classifying Parkinson's disease using structural MRI scans. *Multimedia Tools and Applications*. https://doi.org/10.1007/s11042-023-15302-3

Lakens, D. (2013). Calculating and reporting effect sizes to facilitate cumulative science: A practical primer for t-tests and ANOVAs. *Frontiers in Psychology*, *4*. https://doi.org/10.3389/fpsyg.2013.00863

Liu, L., Wang, Q., Adeli, E., Zhang, L., Zhang, H., & Shen, D. (2016). Feature Selection Based on Iterative Canonical Correlation Analysis for Automatic Diagnosis of Parkinson's Disease. In S. Ourselin, L. Joskowicz, M. R. Sabuncu, G. Unal, & W. Wells (Eds.), *Medical Image Computing and Computer-Assisted Intervention – MICCAI 2016* (pp. 1–8). Springer International Publishing. https://doi.org/10.1007/978-3-319-46723-8_1

Luo, N., Shi, W., Yang, Z., Song, M., & Jiang, T. (2024). Multimodal Fusion of Brain Imaging Data: Methods and Applications. *Machine Intelligence Research*, *21*(1), 136–152. https://doi.org/10.1007/s11633-023-1442-8

Ma, Z., Xu, J., Li, R., & Chen, J. (2023). Gravitational clustering algorithm based on mutual K-nearest neighbors. *Proceedings of the 2023 3rd International Conference on Artificial Intelligence, Automation and Algorithms*, 79–85. https://doi.org/10.1145/3611450.3611462

Newman, M. E. J. (2002). Assortative Mixing in Networks. *Physical Review Letters*, *89*(20), 208701. https://doi.org/10.1103/PhysRevLett.89.208701

Park, C., Lee, P. H., Lee, S.-K., Chung, S. J., & Shin, N.-Y. (2020). The diagnostic potential of multimodal neuroimaging measures in Parkinson's disease and atypical parkinsonism. *Brain and Behavior*, e01808. https://doi.org/10.1002/brb3.1808

Pereira, J. B., Aarsland, D., Ginestet, C. E., Lebedev, A. V., Wahlund, L.-O., Simmons, A., Volpe, G., & Westman, E. (2015). Aberrant cerebral network topology and mild cognitive impairment in early Parkinson's disease: Aberrant Brain Network Topology in Early PD. *Human Brain Mapping*, *36*(8), 2980–2995. https://doi.org/10.1002/hbm.22822

Pfefferbaum, A., Mathalon, D. H., Sullivan, E. V., Rawles, J. M., Zipursky, R. B., & Lim, K. O. (1994). A Quantitative Magnetic Resonance Imaging Study of Changes in Brain

Morphology From Infancy to Late Adulthood. *Archives of Neurology*, *51*(9), 874–887. https://doi.org/10.1001/archneur.1994.00540210046012

Rahaman, M. A., Mathalon, D., Lee, H. J., Jiang, W., Mueller, B. A., Andreassen, O., Agartz, I., Sponheim, S. R., Mayer, A. R., Stephen, J., Jung, R. E., Turner, J. A., Canive, J., Bustillo, J., Calhoun, V. D., Gupta, C. N., Rachakonda, S., Chen, J., Liu, J., … Ford, J. (2020). N-BiC: A Method for Multi-Component and Symptom Biclustering of Structural MRI Data: Application to Schizophrenia. *IEEE Transactions on Biomedical Engineering*, *67*(1), 110–121. https://doi.org/10.1109/TBME.2019.2908815

Rana, B., Juneja, A., Saxena, M., Gudwani, S., Kumaran, S. S., Behari, M., & Agrawal, R. K. (2017). Relevant 3D local binary pattern based features from fused feature descriptor for differential diagnosis of Parkinson's disease using structural MRI. *Biomedical Signal Processing and Control*, *34*, 134–143. https://doi.org/10.1016/j.bspc.2017.01.007

Rana, B., Juneja, A., Saxena, M., Gudwani, S., Senthil Kumaran, S., Agrawal, R. K., & Behari, M. (2015). Regions-of-interest based automated diagnosis of Parkinson's disease using T1-weighted MRI. *Expert Systems with Applications*, *42*(9), 4506–4516. https://doi.org/10.1016/j.eswa.2015.01.062

Rodriguez-Sanchez, F., Rodriguez-Blazquez, C., Bielza, C., Larrañaga, P., Weintraub, D., Martinez-Martin, P., Rizos, A., Schrag, A., & Chaudhuri, K. R. (2021). Identifying Parkinson's disease subtypes with motor and non-motor symptoms via model-based multi-partition clustering. *Scientific Reports*, *11*(1), Article 1. https://doi.org/10.1038/s41598-021-03118-w

Rubinov, M., & Sporns, O. (2010). Complex network measures of brain connectivity: Uses and interpretations. *NeuroImage*, *52*(3), 1059–1069. https://doi.org/10.1016/j.neuroimage.2009.10.003

Salvatore, C., Cerasa, A., Castiglioni, I., Gallivanone, F., Augimeri, A., Lopez, M., Arabia, G., Morelli, M., Gilardi, M. C., & Quattrone, A. (2014). Machine learning on brain MRI data for differential diagnosis of Parkinson's disease and Progressive Supranuclear Palsy. *Journal of Neuroscience Methods*, *222*, 230–237. https://doi.org/10.1016/j.jneumeth.2013.11.016

Samantaray, T., Gupta, U., Saini, J., & Gupta, C. N. (2023). Unique Brain Network Identification Number for Parkinson's and Healthy Individuals Using Structural MRI. *Brain Sciences*, *13*(9), Article 9. https://doi.org/10.3390/brainsci13091297

Samantaray, T., Saini, J., & Gupta, C. N. (2021, March 4). *Meta-Analysis of Clinical Symptoms and Data Driven Subtyping Approaches in Parkinson's Disease* [Poster]. https://thebrainconference.co.uk/wp-content/uploads/2021/02/R2_51_Samantaray_Tanmayee_MovementDisorders_51.png

Samantaray, T., Saini, J., & Gupta, C. N. (2022a). Sparsity Dependent Metrics Depict Alteration of Brain Network Connectivity in Parkinson's Disease. *2022 44th Annual International Conference of the IEEE Engineering in Medicine & Biology Society (EMBC)*, 698–701. https://doi.org/10.1109/EMBC48229.2022.9871258

Samantaray, T., Saini, J., & Gupta, C. N. (2022b). Subgrouping and structural brain connectivity of Parkinson's disease – past studies and future directions. *Neuroscience Informatics*, *2*(4), 100100. https://doi.org/10.1016/j.neuri.2022.100100

Samantaray, T., Saini, J., Pal, P. K., & Gupta, C. N. (2024). Brain connectivity for subtypes of parkinson's disease using structural MRI. *Biomedical Physics & Engineering Express*, *10*(2), 025012. https://doi.org/10.1088/2057-1976/ad1e77

Sardana, D., & Bhatnagar, R. (2014). Graph Clustering Using Mutual K-Nearest Neighbors. In D. Ślęzak, G. Schaefer, S. T. Vuong, & Y.-S. Kim (Eds.), *Active Media Technology* (pp. 35–48). Springer International Publishing. https://doi.org/10.1007/978-3-319-09912-5_4

Schalkamp, A.-K., Rahman, N., Monzón-Sandoval, J., & Sandor, C. (2022). Deep phenotyping for precision medicine in Parkinson's disease. *Disease Models & Mechanisms*, *15*(6), dmm049376. https://doi.org/10.1242/dmm.049376

Shattuck, D. W., Mirza, M., Adisetiyo, V., Hojatkashani, C., Salamon, G., Narr, K. L., Poldrack,

R. A., Bilder, R. M., & Toga, A. W. (2008). Construction of a 3D Probabilistic Atlas of Human Cortical Structures. *NeuroImage*, *39*(3), 1064–1080. https://doi.org/10.1016/j.neuroimage.2007.09.031

Shreeram, A., Samantaray, T., & Gupta, C. N. (2023). Optimizing Permutations in Biclustering Algorithms. In S. Shakya, J. M. R. S. Tavares, A. Fernández-Caballero, & G. Papakostas (Eds.), *Fourth International Conference on Image Processing and Capsule Networks* (pp. 115–129). Springer Nature. https://doi.org/10.1007/978-981-99-7093-3_7

Song, C., Zhao, W., Jiang, H., Liu, X., Duan, Y., Yu, X., Yu, X., Zhang, J., Kui, J., Liu, C., & Tang, Y. (2021). Stability Evaluation of Brain Changes in Parkinson's Disease Based on Machine Learning. *Frontiers in Computational Neuroscience*, *15*. https://www.frontiersin.org/articles/10.3389/fncom.2021.735991

Sui, J., Yu, Q., He, H., Pearlson, G., & Calhoun, V. D. (2012). A Selective Review of Multimodal Fusion Methods in Schizophrenia. *Frontiers in Human Neuroscience*, *6*. https://doi.org/10.3389/fnhum.2012.00027

Sui, J., Zhi, D., & Calhoun, V. D. (2023). Data-driven multimodal fusion: Approaches and applications in psychiatric research. *Psychoradiology*, *3*, kkad026. https://doi.org/10.1093/psyrad/kkad026

Theis, N., Rubin, J., Cape, J., Iyengar, S., & Prasad, K. M. (2023). Threshold Selection for Brain Connectomes. *Brain Connectivity*, *13*(7), 383–393. https://doi.org/10.1089/brain.2022.0082

Wang, Y., Pang, W., & Jiao, Z. (2023). An adaptive mutual K-nearest neighbors clustering algorithm based on maximizing mutual information. *Pattern Recognition*, *137*, 109273. https://doi.org/10.1016/j.patcog.2022.109273

Wu, Q., Gao, Y., Liu, A.-S., Xie, L.-Z., Qian, L., & Yang, X.-G. (2018). Large-scale cortical volume correlation networks reveal disrupted small world patterns in Parkinson's disease. *Neuroscience Letters*, *662*, 374–380. https://doi.org/10.1016/j.neulet.2017.10.032

Xia, M., Wang, J., & He, Y. (2013). BrainNet Viewer: A Network Visualization Tool for Human Brain Connectomics. *PLoS ONE*, *8*(7). https://doi.org/10.1371/journal.pone.0068910

Xu, L., Groth, K. M., Pearlson, G., Schretlen, D. J., & Calhoun, V. D. (2009). Source-based morphometry: The use of independent component analysis to identify gray matter differences with application to schizophrenia. *Human Brain Mapping*, *30*(3), 711–724. https://doi.org/10.1002/hbm.20540

Xu, L., Pearlson, G., & Calhoun, V. D. (2009). Joint source based morphometry identifies linked gray and white matter group differences. *NeuroImage*, *44*(3), 777–789. https://doi.org/10.1016/j.neuroimage.2008.09.051

Xu, X., Guan, X., Guo, T., Zeng, Q., Ye, R., Wang, J., Zhong, J., Xuan, M., Gu, Q., Huang, P., Pu, J., Zhang, B., & Zhang, M. (2018). Brain Atrophy and Reorganization of Structural Network in Parkinson's Disease With Hemiparkinsonism. *Frontiers in Human Neuroscience*, *12*, 117. https://doi.org/10.3389/fnhum.2018.00117

Yadav, S. K., Kathiresan, N., Mohan, S., Vasileiou, G., Singh, A., Kaura, D., Melhem, E. R., Gupta, R. K., Wang, E., Marincola, F. M., Borthakur, A., & Haris, M. (2016). Gender-based analysis of cortical thickness and structural connectivity in Parkinson's disease. *Journal of Neurology*, *263*(11), 2308–2318. https://doi.org/10.1007/s00415-016-8265-2

Zalesky, A., Fornito, A., & Bullmore, E. (2012). On the use of correlation as a measure of network connectivity. *NeuroImage*, *60*(4), 2096–2106. https://doi.org/10.1016/j.neuroimage.2012.02.001

Zeighami, Y., Ulla, M., Iturria-Medina, Y., Dadar, M., Zhang, Y., Larcher, K. M.-H., Fonov, V., Evans, A. C., Collins, D. L., & Dagher, A. (2015). Network structure of brain atrophy in de novo Parkinson's disease. *eLife*, *4*, e08440. https://doi.org/10.7554/eLife.08440

Zhang, D., Wang, J., Liu, X., Chen, J., & Liu, B. (2015). Aberrant Brain Network Efficiency in Parkinson's Disease Patients with Tremor: A Multi-Modality Study. *Frontiers in Aging Neuroscience*, *7*. https://doi.org/10.3389/fnagi.2015.00169